\documentclass[twocolumn,runningheads]{svjour2}
\smartqed  % flush right qed marks, e.g. at end of proof
\usepackage{graphicx}
\usepackage{mathptmx}      % use Times fonts if available on your TeX system
\newcommand{\gr}{$\gamma$-ray \,}

\journalname{Astrophysics and Space Science}
\begin{document}

\title {New evidence for strong nonthermal effects in Tycho's supernova remnant}

\author{H.J. V\"olk \and
        E.G. Berezhko \and
        L.T. Ksenofontov
}
\institute{H.J. V\"olk \at Max Planck Institut f\"ur Kernphysik,
                Postfach 103980, D-69029 Heidelberg, Germany
                \email{Heinrich.Voelk@mpi-hd.mpg.de}  \and 
E.G. Berezhko \and L.T. Ksenofontov \at Yu.G. Shafer Institute of Cosmophysical Research and Aeronomy,
                     31 Lenin Ave., 677980 Yakutsk, Russia}

\date{Received: date / Accepted: date}
% The correct dates will be entered by the editor

\maketitle

\begin{abstract} For the case of Tycho's supernova remnant 
     (SNR) we present the relation between the blast wave and contact
     discontinuity radii calculated within the nonlinear kinetic theory of
     cosmic ray (CR) acceleration in SNRs. It is demonstrated that these radii
     are confirmed by recently published Chandra measurements which show that
     the observed contact discontinuity radius is so close to the shock radius
     that it can only be explained by efficient CR acceleration which in turn
     makes the medium more compressible. Together with the recently determined
     new value $E_\mathrm{sn}=1.2\times 10^{51}$~erg of the SN explosion energy
     this also confirms our previous conclusion that a TeV \gr flux of
     $(2-5)\times 10^{-13}$~erg/(cm$^2$s) is to be expected from Tycho's
     SNR. {\it Chandra} measurements and the {\it HEGRA} upper limit of the TeV
     \gr flux together limit the source distance $d$ to $3.3\leq d\leq 4$~kpc.
\end{abstract}

%________________________________________________________________

\section{Introduction}

Cosmic rays (CRs) are widely accepted to be produced in SNRs by the diffusive
shock acceleration process at the outer blast wave (see e.g. 
\cite{drury,be87,bk88,jel91,mald01} for
reviews). Kinetic nonlinear theory of diffusive
CR acceleration in SNRs \cite{byk96,bv97} couples the gas dynamics of the
explosion with the particle acceleration. Therefore in a spherically symmetric
approach it is able to predict the evolution of gas
density, pressure, mass velocity, as well as the positions of the forward shock
and the contact discontinuity, together with the energy spectrum and the
spatial distribution of CR nuclei and electrons at any given evolutionary epoch
$t$, including the properties of the nonthermal radiation. 
The application of this theory
to individual SNRs \cite{bkv02,bkv03,bpv03,vbkr} has demonstrated its power in
explaining the observed SNR properties and in predicting new effects like the
extent of magnetic field amplification, leading to the concentration of the
highest-energy electrons in a very thin shell just behind the shock.

Recent observations with the {\it Chandra} and {\it XMM-Newton} X-ray
telescopes in space have confirmed earlier detections of nonthermal continuum
emission in hard X-rays from young shell-type SNRs. With {\it Chandra} it
became even possible to resolve spatial scales down to the arcsec extension of
individual dynamical structures like shocks \cite{vl03,long,bamba}. The
filamentary hard X-ray structures are the result of strong synchrotron losses
of the emitting multi-TeV electrons in amplified magnetic fields downstream of
the outer accelerating SNR shock \cite{vl03,bkv03,bv04b,vbk05}. Such
observational results gain their qualitative significance through the fact that
these effective magnetic fields and morphologies turned out to be exactly the
same as predicted theoretically from acceleration theory.
 
This theory has been applied in detail to Tycho's SNR, in order to compare
results with the existing data \cite{vbkr,vbk05}.  We have used a stellar
ejecta mass $M_{ej}=1.4M_{\odot}$, distance $d=2.3$~kpc, and interstellar
medium (ISM) number density $N_H=0.5$~ H-atoms cm$^{-3}$.  For these parameters
a total hydrodynamic explosion energy $E_{sn}=0.27\times 10^{51}$~erg was
derived to fit the observed size $R_s$ and expansion speed $V_s$.  A rather
high downstream magnetic field strength $B_d\approx 300$~$\mu$G and a proton
injection rate $\eta=3\times 10^{-4}$ are needed to reproduce the observed
steep and concave radio spectrum and to ensure a smooth cutoff of the
synchrotron emission in the X-ray region. We believe that the required strength
of the magnetic field, that is significantly higher than the 
MHD compression of a $5~\mu$G ISM field, has to be
attributed to nonlinear field amplification at the SN shock by CR acceleration
itself.  According to plasma physical considerations
\cite{lb00,belll01,bell04}, the existing ISM magnetic field can indeed be
significantly amplified at a strong shock by CR streaming instabilities.

After adjustment of the predictions of the nonlinear sphe\-ri\-cally-symmetric
model by a physically necessary renormalization of the number of accelerated CR
nuclei to take account of the quasi-perpendicular shock directions in a SNR,
very good consistency with the existing observational data was achieved.

Using {\it Chandra} X-ray observations \cite{warren05} have recently
estimated the ratio between the radius
$R_\mathrm{c}$ of the contact discontinuity (CD), separating the swept-up ISM
and the ejecta material, and the radius $R_\mathrm{s}$ of the forward
shock. The large mean value $R_\mathrm{c}/R_\mathrm{s}=0.93$ of this ratio was
interpreted as evidence for efficient CR acceleration, which makes the medium
between those two discontinuities more compressible.

Here we present the calculations of the mean ratio $R_\mathrm{c}/R_\mathrm{s}$,
which are the unchanged part of our earlier considerations \cite{vbkr,vbk05},
and demonstrate that these results, which are in fact predictions, fit the
above measurements very well.  Since our calculations have been made in
spherical symmetry they concern a priori an azimuthally averaged ratio
$R_\mathrm{c}/R_\mathrm{s}$. We shall extend them by taking the effects of the
Rayleigh-Taylor (R-T) instability of the CD into account. We shall in addition
discuss a physical mechanism that leads to the observed {\it azimuthal}
variations of $R_\mathrm{c}/R_\mathrm{s}$. Finally we shall take the recent
determination of the mechanical energy output $E_\mathrm{sn}\approx 1.2 \times
10^{51}$ \cite{bad05} which results from the theory of delayed detonations in
the physics of type Ia SN explosions \cite{gam04,gam05}, in order to predict a
range for the \gr spectrum that is consistent with the existing upper limits of
high energy \gr fluxes from Tycho's SNR.

\section{Results and Discussion}

Fig.1 and partly Fig.2 show the calculations of shock and CD
related quantities which were part of our earlier considerations
\cite{vbkr,vbk05}. The calculated shock as well as CD radii and speeds are
shown as a function of time for the two different cases of interior magnetic
field strengths $B_\mathrm{d}=240$~$\mu$G and $B_\mathrm{d}=360$~$\mu$G
considered, together with the azimuthally averaged experimental data available
at the time.
%------------------------------------------------------------------------fig.1-
\begin{figure}
\centering
\includegraphics[width=0.45\textwidth]{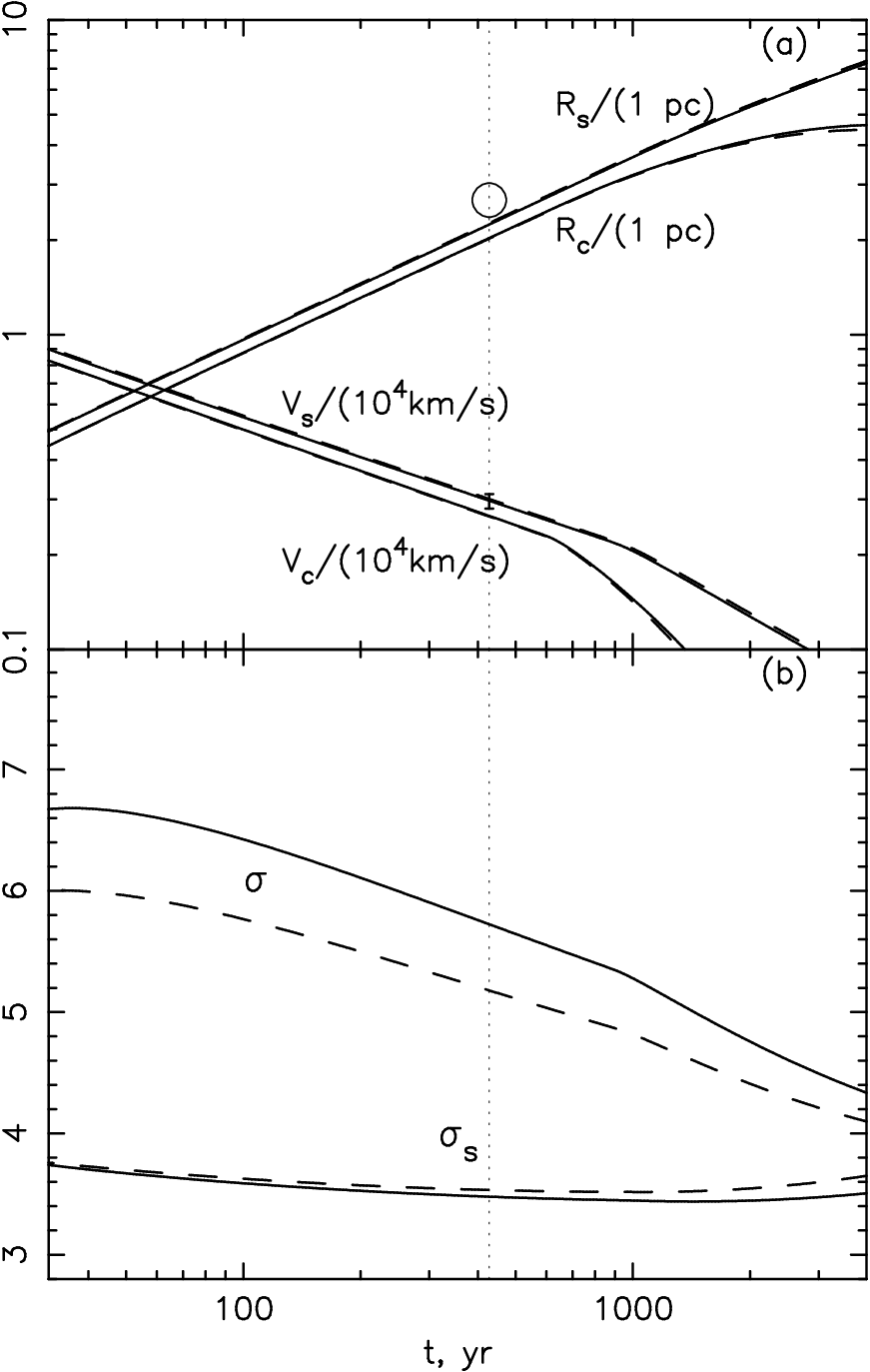}
\caption{(a) Shock radius $R_\mathrm{s}$, contact
discontinuity radius $R_\mathrm{c}$, shock speed $V_\mathrm{s}$, and contact
discontinuity speed $V_\mathrm{c}$, for Tycho`s SNR as functions of time,
including particle acceleration; 
(b) total shock ($\sigma$) and subshock
($\sigma_\mathrm{s}$) compression ratios. The {\it dotted vertical line} marks
the current epoch. The {\it solid and dashed lines} correspond to the internal
magnetic field strength $B_\mathrm{d}=240$~$\mu$G and
($B_\mathrm{d}=360$~$\mu$G), respectively.  The observed mean size and speed of
the shock, as determined by radio measurements
\cite{tg85}, are shown as well.}
\label{f1}
\end{figure}
%------------------------------------------------------------------------------

According to Fig.1a Tycho is nearing the adiabatic phase. To fit the spectral
shape of the observed radio emission we assumed a proton
injection rate $\eta=3\times 10^{-4}$. This leads to a significant nonlinear
modification of the shock at the current age of $t=428$~yrs. A larger magnetic
field lowers the Alfv\'enic Mach number and therefore leads to a decrease of
the shock compression ratio, as seen in Fig.1b. The result is a total
compression ratio $\sigma=5.7$ and a subshock compression ratio $\sigma_s=3.5$
for $B_\mathrm{d}=240$~$\mu$G. In turn $\sigma=5.2$, $\sigma_s=3.6$, for
$B_\mathrm{d}=360$~$\mu$G.

Therefore, as can be seen from Fig.2, including CR acceleration at the outer
blast wave, the calculated value of the ratio $R_\mathrm{c}/R_\mathrm{s}$ for
$B_\mathrm{d}=360$~$\mu$G is slightly lower than for $B_\mathrm{d}=240$~$\mu$G.
At the current epoch we have $R_\mathrm{c}/R_\mathrm{s}\approx 0.90$ which is
lower than the value $R_\mathrm{c}/R_\mathrm{s}=0.93$ inferred from the
observations. Qualitatively our result goes in the same direction as
calculations by \cite{blon01} which modeled SNRs with a uniform specific heat
ratio $\gamma_\mathrm{eff} < 5/3$, for the circumstellar medium and the ejecta
material alike.

Projecting a highly structured shell onto the plane of the sky tends to favor
protruding parts of the shell. Therefore the average radius measured in
projection is an overestimate of the true average radius. Analysing the amount
of bias from the projection for the shock and CD radii \cite{warren05} found a
corrected "true" value $R_\mathrm{c}/R_\mathrm{s}=0.93$ which is lower than
their measured ``projected average'' value $R_\mathrm{c}/R_\mathrm{s}=0.96$, as
a result of the above geometrical effect.

In turn, starting from a spherically symmetric calculation of the CD radius, as
we do, one has to take into account that the actual CD is subject to the R-T
instability. In the nonlinear regime it leads to effective mixing of the ejecta
and swept-up ISM material with ``fingers'' of the ejecta on top of this mixing
region, which extend farther into the shocked gas than the radius
$R_\mathrm{c}$ predicted when assuming spherical symmetry
e.g. \cite{chevetal92,dwark00,blon01,wang01}. Therefore our ratio
$R_\mathrm{c}/R_\mathrm{s}=0.90$, calculated within the spherically symmetric
approach, has to be corrected for this effect in order to compare it with the
measured value $R_\mathrm{c}/R_\mathrm{s}=0.93$.  In the case when all the
fingers have length $l$ and occupy half of the CD surface, one would have a
mean CD size $R_c'\approx R_c+0.5l$ which has to be compared with
$0.93R_\mathrm{s}$.  According to the numerical modelling of \cite{wang01},
albeit without particle acceleration, the R-T instability allows fingers of
ejecta to protrude beyond the spherically symmetric CD radius by 10\%.  The
longest fingers of size $l\approx 0.1R_c$ occupy less than 50\% of the CD
surface.  However, in projection they stick out of the mixing region, whose
thickness is roughly $0.5l$.  This leads to a rough estimate of the corrected
CD radius $R'_c = 1.05R_c$ which has to be compared with the experimentally
estimated value.

The comparison of the corrected values $R'_\mathrm{c}/R_\mathrm{s}$, according
to our earlier calculations as well as for different assumptions (see below)
about the explosion energy and source distance, with this experimentally
estimated value $R_\mathrm{c}/R_\mathrm{s}=0.93$ (in Fig.2 we present that
value with 2\% uncertainties, according to \cite{warren05}) shows quite good
agreement (see Figs.2 and 3) even if one takes into account some uncertainty in
the quantitative determination of our correction factor
$R'_\mathrm{c}/R_\mathrm{c}$, which in our view lies in the range 1.03-1.07.
% 
%------------------------------------------------------------------------fig.2-
\begin{figure} 
\centering
\includegraphics[width=0.45\textwidth]{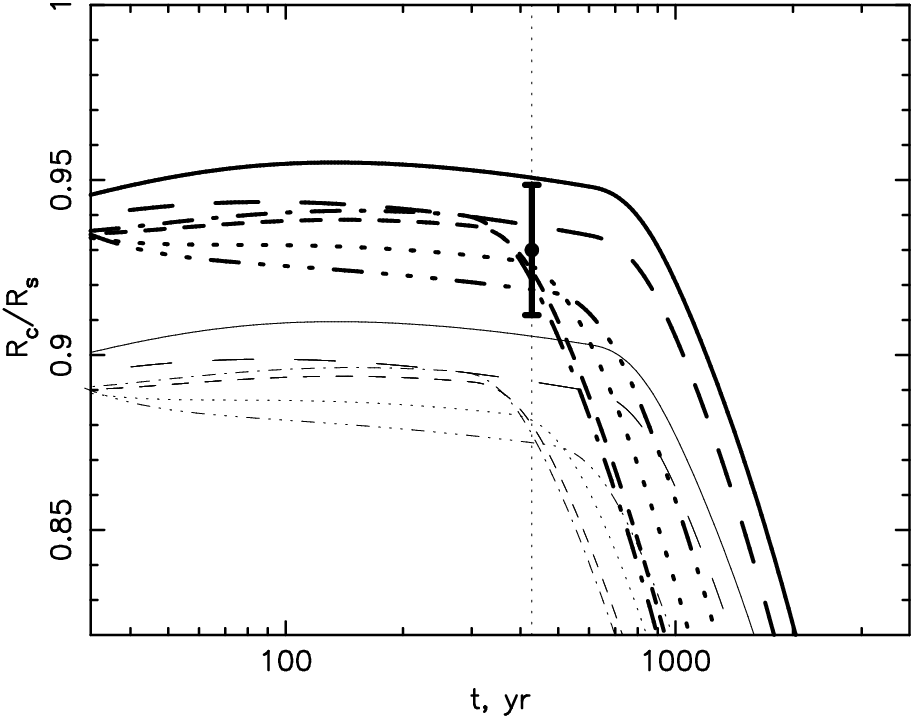}
\caption{
The ratio $R_\mathrm{c}/R_\mathrm{s}$ of the radii of the contact discontinuity
and the forward shock  as a function of
time. {\it Solid and dashed} lines correspond to the same two cases as in
Fig.1. The lines of all other styles correspond to the SN explosion energy
$E_\mathrm{sn}=1.2\times 10^{51}$~erg and four different distances (see the
caption of Fig.3) {\it Thin} lines represent the values calculated in the
spherically symmetric model, whereas the {\it thick} lines show the values
$R'_\mathrm{c}/R_\mathrm{s}$ which contain the correction for the effect
produced by the R-T instability. The experimental point is taken
from \cite{warren05}.}
\label{f2} \end{figure}
%------------------------------------------------------------------------------
%------------------------------------------------------------------------fig.3-
\begin{figure} 
\centering
\includegraphics[width=0.45\textwidth]{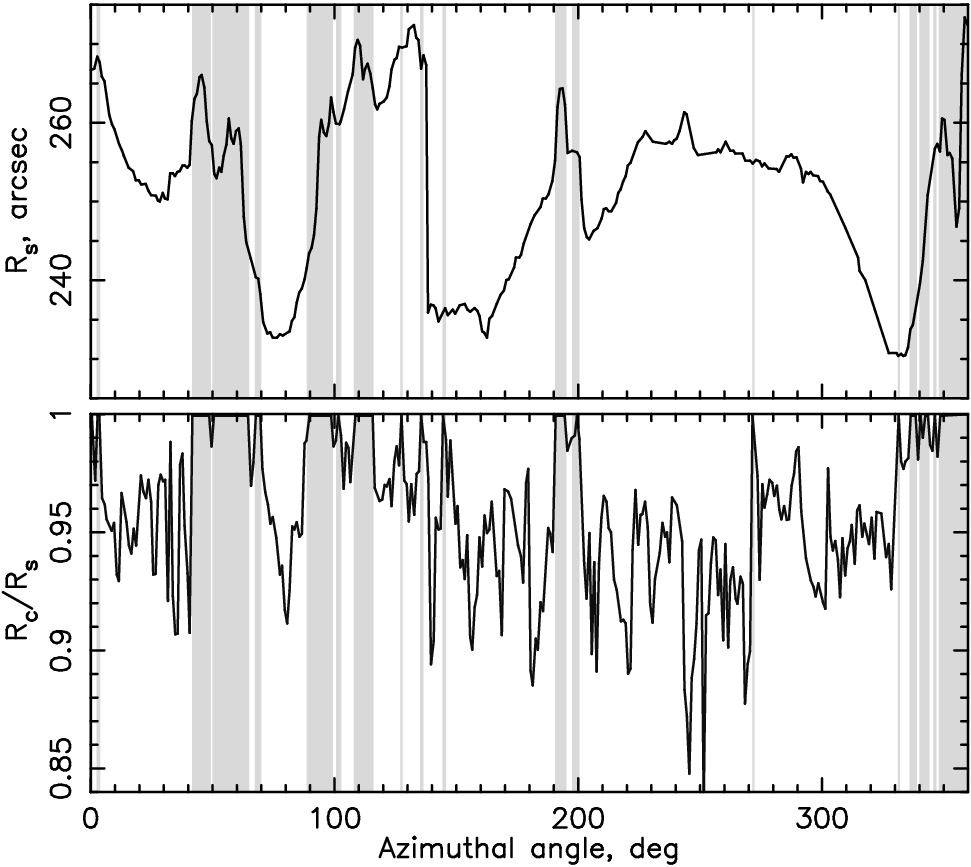}
\caption{{\em Top panel}: the forward shock radius $R_\mathrm{s}$ and
{\em bottom panel}: the ratio $R_\mathrm{c}/R_\mathrm{s}$ of the radii 
of the contact discontinuity
and the forward shock as a function of azimutal angle \cite{warren05}.
The regions where $R_\mathrm{c}/R_\mathrm{s}> 0.99$ are shadowed.
}
\label{w} \end{figure}
%------------------------------------------------------------------------------

Another interesting peculiarity of Tycho's shock structure, which we would like
to discuss here, is the quite irregular behaviour of the radius of the forward
shock around the edge of the visible SNR disk, that is clearly seen in
Fig.\ref{w}. Large shock distortions of this kind are
not expected to result from the R-T instability. At first sight such a
variation of the values of $R_\mathrm{s}$ and $R_\mathrm{c}$ can be easily
attributed to fluctuations of the ambient ISM density and/or to an
inhomogeneously distributed density and velocity field of the ejected
matter. The local shock part, which encounters a lower ISM density or has a
faster ejecta portion behind, propagates faster compared with the neighbouring
shock pie\-ces. This will lead to the formation of local forward displacements of
the CD and the forward shock whose number and relative sizes are determined by
the specific structure of the ISM and/or ejecta. Since a larger shock
compression ratio is expected for higher shock speed (see Fig.1b), one should
also expect a smaller difference between $R_\mathrm{s}$ and $R_\mathrm{c}$ on
the top region of each such displacement. This is exactly what is observed.

Such arguments would give a good explanation for the observed picture only if
CR injection/acceleration took place uniformly across the entire shock
surface. However, in our view the actual situation is expected to be more
complicated -- and physically more interesting. Efficient injection of
suprathermal nuclear particles into the acceleration process takes place in
those local shock regions, where the forward shock is quasi-parallel, and these
regions are distributed over the shock surface according to the ambient ISM
magnetic field structure and occupy in total about 20\% of the shock
\cite{vbk03}. As a result one would expect that only about 20\% of the local
forward shock displacements efficiently accelerate nuclear CRs, and therefore
display the extraordinary high ratio $R_\mathrm{c}/R_\mathrm{s}$. In reality
the most extreme values $R_\mathrm{c}/R_\mathrm{s}\geq 0.99$ are observed on
randomly distributed local shock regions, and the positions of these regions
roughly coincide with the positions of extended shock displacements. To explain
such a correlation a strong physical connection between the shock speed and the
efficiency of CR injection/acceleration should exist: the local parts of the
shock with large speed effectively produce CRs and vice versa.

There are at least two physical processes which can resolve the above
problem. The first one gives high injection on the leading part of the
displacements, if they are formed due to ISM and/or ejecta inhomogeneities and
initially the injection was suppressed here as a result of a highly oblique
magnetic field.  Since after its formation the leading part of the displacement
develops a large curvature, it will have a significant portion which becomes
quasi-parallel, and therefore efficient CR injection/acceleration is expected
over the whole leading part of such a displacement.

The second factor can presumably itself lead to the formation of displacements,
even in the case of a uniform ISM and a spherically symmetric ejecta
distribution. A significant fraction of the internal energy behind a
quasi-parallel part of the shock front is contained in CRs. Since the shock is
expected to be modified, the CR spectrum is very hard and therefore the
pressure is mainly carried by the CRs with the highest energies. With their
high mobility these CRs can diffuse laterally into neighbouring downstream
volumes, which are located behind shock surfaces not producing CRs. This
diffusive loss leads to a decrease of the internal pressure behind the
quasi-parallel parts of the shock. Therefore the corresponding downstream
ejecta undergo less deceleration. As a result these ejecta accelerate relative
to their surroundings. This leads to the formation of a local displacement. One
can expect that this also distorts the initially spherically symmetric
shock. We therefore conclude that the effectively accelerating parts of the
forward shock are among those regions, where the displacements occur, and they
may actually delineate them. Predominantly fingers are situated on the
effectively accelerating parts of the shock surface. If this is correct it
gives a unique experimental identification of the areas at the outer SNR shock
which efficiently produce nuclear CRs.

Since the most outwardly displaced parts of the CD and the shock with a high
ratio $R_\mathrm{c}/R_\mathrm{s}$ have to be interpreted as the regions with
efficient CR injection/acceleration, this opens the possibility to
experimentally distinguish the shock areas with efficient CR production from
those where CRs are not produced. A rough estimate shows that in Tycho's SNR
the regions with extremely high ratios $R_\mathrm{c}/R_\mathrm{s}>0.99$ and the
displacements occupy about 20\% of the shock surface see Fig.4 of \cite{warren05}.
This corresponds to the theoretical expectation.

Our last point regards constraints which the recent re-evaluation of the
mechanical output $E_\mathrm{sn}=1.2 \times 10^{51}$~erg together with the {\it
HEGRA} upper limit for the TeV \gr flux \cite{aha01} approximately
impose on the distance and ambient density for Tycho's SNR. With this new
$E_\mathrm{sn}$-value we find a consistent fit for all existing data -- the SNR
size, its expansion rate, overall synchrotron spectrum and the filament
structure of the X-ray emission -- like it was done for the previously defined
explosion energy $E_\mathrm{sn}=0.27 \times10^{51}$~erg (see
\cite{vbkr,vbk05} for details). In particular, the fit of $R_\mathrm{s}$ and
$V_\mathrm{s}$, which is of the same quality as in Fig.1 for
$E_\mathrm{sn}=0.27 \times10^{51}$~erg, gives for each assumed distance $d$ a
rather definite value of the ISM number density $N_\mathrm{H}(d)$. A rough
explanation of this numerical result is the following: since for a given
experimentally measured angular SNR size and its expansion rate the linear size
$R_\mathrm{s}$ and the speed $V_\mathrm{s}$ scale proportionally to distance
$d$, and since in the nearby (in time) Sedov phase $R_\mathrm{s}\propto
(E_\mathrm{sn}/N_\mathrm{H})^{1/5}$, the density $N_\mathrm{H}\propto
E_\mathrm{sn}/d^5$ decreases with increasing distance $d$. The hadronic \gr
flux $F_{\gamma}\propto R_\mathrm{s}^3V_\mathrm{s}^2 N_\mathrm{H}^2/d^2$ is
then expected to scale as $F_{\gamma}\propto E_\mathrm{sn}^2/d^7$.  The ejected
mass is still assumed to be $M_{\mathrm{ej}}=1.4 M_{\odot}$. We also find the
same nuclear injection rate $\eta=3\times 10^{-4}$ for all cases, and
downstream magnetic field values $B_\mathrm{d}\approx 400$~$\mu$G. At the same
time, the linear size $L$ of an X-ray filament increases proportional to
$d$. Therefore the magnetic field strength $B'_\mathrm{d}\propto L^{-2/3}$
\cite{bv04b}, determined from the filament sizes (see Introduction), decreases
with $d$.
% 
%------------------------------------------------------------------------fig.4-
\begin{figure} 
\centering
\includegraphics[width=0.45\textwidth]{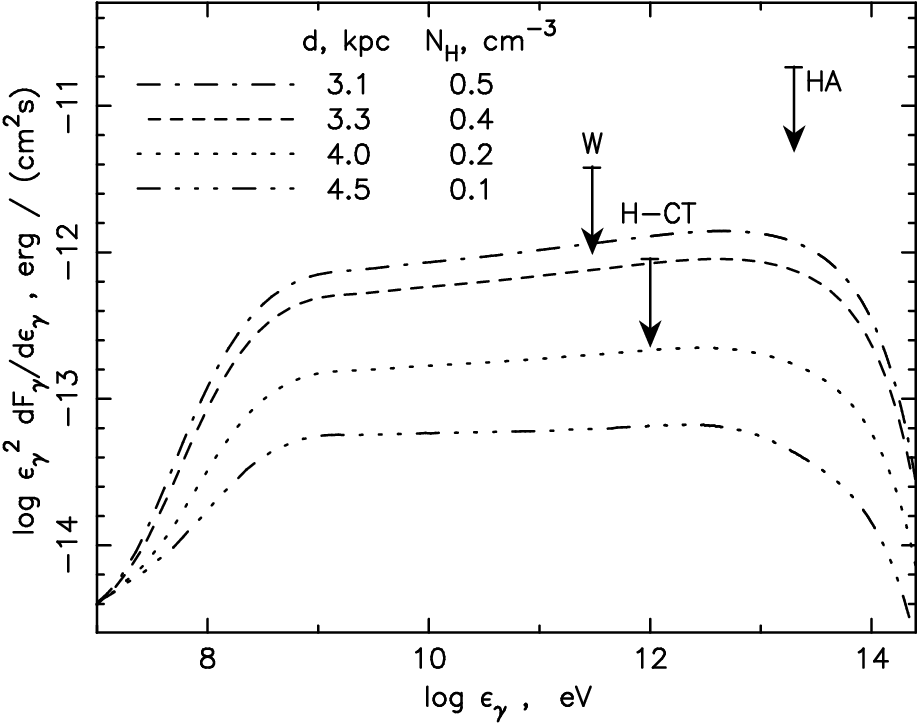}
\caption{
Spectral energy distribution of the \gr emission from Tycho's SNR, as a
function of \gr energy $\epsilon_{\gamma}$, for a mechanical SN explosion
energy of $E_\mathrm{sn}= 1.2 \times 10^{51}$~erg and four different distances
$d$ and corresponding values of the ISM number densities $N_\mathrm{H}$. All
cases have a dominant hadronic compared to Inverse Compton \gr
flux. Experimental data are the upper limits of the {\it HEGRA} (H-CT;
\cite{aha01}) and {\it Whipple} (W; \cite{buckley98}) Cherenkov
telescopes and the 95\% confidence {\it HEGRA AIROBICC} (HA; \cite{prahl97})
upper limit.}
\label{f3} 
\end{figure}
%------------------------------------------------------------------------------
%

In order to find the constraint on the distance $d$ and the ISM density
$N_\mathrm{H}$, we then compare in Fig.3 the resulting \gr spectral energy
distribution with the {\it HEGRA} and Whipple upper limits at TeV energies. It
is seen that all distances $d < 3.3$~kpc are inconsistent with the {\it HEGRA}
data. Distances of $d>4$~kpc are still consistent with the \gr data, although
there is a growing discrepancy between $B_\mathrm{d}$ and $B'_\mathrm{d}$: at
$d=4.5$~kpc $B'_\mathrm{d}\approx 300$~$\mu$G which is already considerably
smaller than $B_\mathrm{d}\approx 400$~$\mu$G. Therefore we believe that we can
constrain the source distance also from above, $d<4$~kpc.  In Fig.2 we also
show the values for $R_\mathrm{c}/R_\mathrm{s}$ and
$R'_\mathrm{c}/R_\mathrm{s}$ for the case $E_\mathrm{sn}=1.2 \times
10^{51}$~erg and these increased distances. Within our approximate
determination of $R'_\mathrm{c}$ from $R_\mathrm{c}$ they still agree with the
{\it Chandra} data, in particular because the CR production rates are
comparable. Our calculations of the \gr emission lead us to predict that the
new Northern Hemisphere TeV detectors should detect this source at TeV-energies
in, predominantly, hadronic $\gamma$-rays: the expected $\pi^0$-decay
$\gamma$-ray energy flux $(2-5)\times 10^{-13}$~erg/(cm$^2$s) extends up to
almost 100~TeV if the distance is indeed within the range 3.3-4 kpc. As a
corollary the detection of a TeV signal is not only important by itself, but it
is also crucial for the correct determination of all other key Supernova
parameters.

\begin{acknowledgements}
EGB and LTK acknowledge the partial support by the Presidium of RAS
(program No.16) and by the SB RAS (CIP-2006 No.3.10) and the
hospitality of the Max-Planck-Institut f\"ur Kernphysik, where part of
this work was carried out.
\end{acknowledgements}

\end{document}